\documentclass[11pt]{scrartcl}

\usepackage[utf8]{inputenc}
\usepackage[T1]{fontenc}
\usepackage[english]{babel}
\usepackage{bbold}
\usepackage{lmodern}
\usepackage{microtype}
\usepackage{bm}

\usepackage{amsmath,amssymb,amsfonts}
\usepackage{bm}

\usepackage{graphicx}
\usepackage{hyperref}

\usepackage{geometry}
\usepackage{float}
\usepackage{xcolor}

\usepackage[format=plain,
            singlelinecheck=false,
            indention=0em,
            labelfont=bf]{caption}

\usepackage[symbol]{footmisc}

\newcommand{\email}[1]{\footnote{\texttt{#1}}}

\begin{document}

\begin{center}
    {\Large \textbf{Linear stability of charged warm holes} \\[1.2em]}
    {Radouane Gannouji\email{radouane.gannouji@pucv.cl}},
    {Alejandro Guzmán
Antonucci\email{alejandro.guzman.a@pucv.cl} \\[0.9em]}
{\small Instituto de Física, Pontificia Universidad Católica de Valparaíso Av. Brasil 2950, Valparaíso, Chile.}
\end{center}

\vspace{1em}

\begin{abstract}
Charged black holes are known to suffer from an interior instability associated with the presence of the Cauchy horizon. Recently, a hairy charged black hole was proposed that avoids the formation of a Cauchy horizon. It is natural to question whether this instability might manifest in the exterior solution.  In this paper, we have analyzed the stability of this black hole. Our results show that vector perturbations are stable, along with the scalar sector for $l=0$. We have also computed the corresponding quasinormal modes (QNMs) and quasibound states (QBSs). 
\end{abstract}

\section{Introduction}

Black holes have been extensively studied over the past few decades, with particular attention given to their exterior solutions, as this is the only region accessible to an external observer. By contrast, the interior region of a black hole has received less attention, even though it holds the limits of our current gravitational theory. Any advancement in understanding general relativity must focus on the interior structure. Notably, the interiors of Reissner-Nordström and Kerr black holes feature a Cauchy horizon (CH), which marks the boundary of predictability in classical theory.
The existence of the Cauchy horizon inspired Penrose to propose the Strong Cosmic Censorship conjecture (SCC). This conjecture asserts that, as the Cauchy horizon separates a globally hyperbolic region from a non-hyperbolic one, it should not form. In other words, the conjecture posits that the predictable region of spacetime cannot be meaningfully extended beyond a certain point. If the maximal Cauchy development is inextendible, spacetime ends naturally without introducing "new" physics or structures beyond what can be deduced from initial conditions, meaning no Cauchy horizon should form.
When a Cauchy horizon forms, it is expected to produce an interesting phenomenon known as mass inflation \cite{Poisson:1989zz,Poisson:1990eh,Ori:1991zz}. In such cases, the Cauchy horizon is replaced by a singularity. Early analyses revealed that perturbations might extend continuously across this singularity, although their derivatives would not ($\mathcal{C}^0$ regularity). Consequently, while curvature could diverge, some geodesics might still extend continuously, leading to the concept of weak null singularities. This scenario has been supported by numerous studies, particularly in \cite{Dafermos:2003wr,Dafermos:2003vim}, where it was shown that under conditions like spherical symmetry, curvature diverges, but the metric can extend across the weak null singularity.
In the modern formulation due to Christodoulou, SCC demands inextendibility of the maximal development as a \emph{weak} ($H^1_{\mathrm  loc}$) solution across the CH. In practice, for spacetimes with an inner horizon the relevant competition is between exterior late–time decay and the blueshift at the CH. A precise diagnostic uses the spectral gap of linear perturbations \cite{Cardoso:2017soq}
\begin{align}
\beta \equiv -\frac{\mathrm{Im}\,\omega_0}{\kappa_-}\,,
\end{align}
with $\omega_0$ the least–damped quasinormal frequency and $\kappa_-$ the CH surface gravity. 
If $\beta>1/2$, first derivatives remain square–integrable at the CH and weak extensions exist (violation of modern SCC); if $\beta<1/2$, the $H^1$ energy blows up and SCC holds \cite{Cardoso:2017soq,Dias:2018etb}. For asymptotically de Sitter geometries, a broad body of work shows that near extremality one can have $\beta>1/2$ and hence weakly regular CHs, i.e.\ violations of SCC in this modern sense, in Reissner–Nordström–de~Sitter and closely related setups with neutral/charged matter, higher dimensions, or non-minimal couplings \cite{Cardoso:2018nvb,Mo:2018nnu,Dias:2018ufh,Ge:2018vjq,Liu:2019lon,Liu:2019rbq,Guo:2019tjy,Destounis:2018qnb,Destounis:2019omd,Destounis:2020yav}. Rough initial data \cite{Dafermos:2018tha} considerations and the sensitivity of the quasinormal spectrum further refine this picture \cite{Dias:2018etb,Courty:2023rxk}. See also the recent note \cite{Chrysostomou:2025qud} for a concise overview focused on the RNdS context.

Similarly to the historical debate on the singularity of black holes, as demonstrated in the spherical case by Oppenheimer and Snyder, the question arose whether such singularities are generic or merely a consequence of the unphysical symmetry of the situation. Naturally, this led to the question of whether weak null singularities can exist in spacetimes that are not spherically symmetric. 
Unfortunately, it was demonstrated that weak null singularities are generic \cite{Luk:2013cqa}. Thus, a mechanism preventing Cauchy horizon formation is very welcome, leading to an interior resembling the Schwarzschild case, characterized by a spacelike singularity without Cauchy horizon. The expectation is that, due to some yet unknown mechanism, the Cauchy horizon does not form, and the interior region instead resembles that of the Schwarzschild solution, featuring a spacelike singularity. This naturally raises the question of what mechanism could give rise to such an interior structure.

Regarding whether weak null singularities are generic or an artifact of symmetry, it has been shown that allowing \emph{rough} initial data at the Cauchy horizon heals modern (Christodoulou) SCC even for near-extremal de Sitter black holes, without conflicting with the weak–solution formulation \cite{Dafermos:2018tha}. Concerning non-spherical cases, Kerr–de Sitter \emph{respects} SCC for all non-extremal parameters \cite{Dias:2018ynt}, whereas Kerr–Newman–de Sitter admits near-extremal regions where SCC is \emph{violated} \cite{Casals:2020uxa}.

Recently, attention has turned to a class of charged, hairy black holes without a Cauchy horizon \cite{Cai:2020wrp,Devecioglu:2021xug,Grandi:2021ajl,An:2021plu,Mansoori:2021wxf,Dias:2021afz}. This idea originates from the AdS/CFT correspondence, where deforming the dual CFT with a scalar operator eliminates the Cauchy horizon in the bulk \cite{Hartnoll:2020rwq,Hartnoll:2020fhc}. In flat spacetimes, these hairy black holes not only eliminate the Cauchy horizon but also exhibit intriguing properties, such as extremal solutions with nonzero temperature, referred to as "warm holes" \cite{Dias:2021vve}. This is notable because extremal black holes were previously thought to violate the third law of thermodynamics. Recent studies suggest processes that could form such objects \cite{Kehle:2023eni,Kehle:2024vyt}, implying a possible violation of the third law. However, these extremal hairy black holes, even if formed from generic initial conditions, avoid the violation of the third law, because their temperature is non zero.

Of course to be considered viable solutions to the interior's unpredictability, these configurations must be stable. It has been suggested that proving the stability of Kerr exterior solutions could imply the stability of the Cauchy horizon as a weak null singularity \cite{Dafermos:2017dbw}. Conversely, one might ask whether instability of the Cauchy horizon implies instability of the exterior. In this context, the goal of this paper is to investigate whether the exteriors of these solutions could be unstable. Specifically, we analyze the linear stability of the charged warm hole proposed in \cite{Dias:2021vve}.

Beyond their theoretical role for SCC (the least–damped QNM enters the spectral–gap diagnostic $\beta$ above), linear perturbations also control the observable ringdown. Quasinormal modes (QNMs) are defined by purely ingoing behavior at the horizon and purely outgoing behavior at infinity; they form the backbone of black-hole spectroscopy and parameter estimation with LIGO–Virgo–KAGRA \cite{Kokkotas:1999bd,Berti:2009kk,Konoplya:2011qq,Carullo:2025oms,Berti:2025hly}. Recent work has also clarified caveats such as non-normal effects and spectral sensitivity \cite{Cheung:2021bol,Boyanov:2022ark,Destounis:2023ruj}. By contrast, quasibound states (QBSs) are modes that decay at infinity and remain localized, arising for massive/charged fields or in horizonless settings, and they can dominate late-time dynamics or trigger instabilities \cite{Furuhashi:2004jk,Dolan:2007mj,Hod:2015goa,Huang:2020pga}. QBSs are likewise central in analog-gravity platforms, where their relation to QNMs underpins “analog spectroscopy” \cite{Vieira:2021xqw,Vieira:2021ozg,Vieira:2023ylz,Vieira:2025ljl,Vieira:2025gcy}. Both spectra are discrete sets of complex frequencies whose imaginary parts govern decay or growth; we therefore compute QNMs and QBSs for the warm-hole backgrounds below.

The paper begins with an introduction to the model and background solution, followed by an analysis of scalar and vector perturbations. Finally, we compute the quasinormal modes (QNM) and quasibound states (QBS) before presenting our conclusions.

\section{Equations of motion}

The theory described in \cite{Dias:2021vve} is an Einstein-Maxwell-charged scalar field theory with a complex scalar field coupled to the Maxwell field. The action is the following
\begin{align}
\label{action}
S = \int d^4x \sqrt{-g}\Bigl[& R- F^2 - 4(\mathcal{D}_\mu \psi)(\mathcal{D}^\mu \psi)^\dagger-4m^2|\psi|^2-4\kappa F^2|\psi|^2\Bigr],
\end{align}

\noindent
where $R$ is the Ricci scalar, $\mathcal{D}_\mu = \nabla_\mu - ie A_\mu$ with $e$ being the electric charge and $m$ the mass of the scalar field. On the other hand $F^2 = F_{\mu\nu}F^{\mu\nu}$, where the Maxwell field's $A_\mu$ strength is defined as $F_{\mu\nu} = \partial_\mu A_\nu-\partial_\nu A_\mu$. Lastly, $\kappa$ is the positive coupling constant between the scalar and vector fields.\\
Assuming a static spherically symmetric spacetime 
\begin{align}
\label{line}
ds^2 &=  \Bar{g}_{\mu\nu}dx^\mu dx^\nu=  -A(r) dt^2 + \dfrac{dr^2}{B(r)} + r^2(d\theta^2 + \sin^2{(\theta)} d\phi^2).
\end{align}
The Maxwell and scalar fields, $\Bar{A}_\mu,\Bar{\psi}$ take the following form
\begin{align} 
\label{back_fields}
\Bar{A} = \Bar{A}(r) dt, \qquad \Bar{\psi} =  \Bar{\psi}^\dagger = \Bar{\psi}(r),
\end{align}
where the scalar field, $\Bar{\psi}$, has been chosen to be real. By replacing the background metric, the scalar and vector fields in (\ref{action}) and varying with respect to the functions $X = \{A, B, \Bar{\psi}, \Bar{A}\}$, we obtain the equations of motion (EOM) of the model. Expressing these in the form $\mathcal{E}_X=0$, we obtain
\begin{align}
\label{eqA}
\mathcal{E}_A &\equiv  \frac{1}{A\sqrt{AB}}\Bigl[A(-1 + B + 2m^2r^2\Bar{\psi}^2 + 2r^2B \Bar{\psi}'^2)+ rAB' + r^2(B \Bar{A}'^2 + 2\Bar{\psi}^2(e^2\Bar{A}^2 + 2\kappa B \Bar{A}'^2))\Bigr]\,,\\
\label{eqB}
\mathcal{E}_B &\equiv  \frac{1}{B\sqrt{AB}}\Bigl[A(-1 + B + 2m^2r^2\Bar{\psi}^2 - 2r^2B \Bar{\psi}'^2) rBA' + r^2(B \Bar{A}'^2 - 2\Bar{\psi}^2(e^2\Bar{A}^2 - 2\kappa B \Bar{A}'^2))\Bigr]\,,\\
\label{eqPsi}
\mathcal{E}_{\Bar{\psi}} &\equiv  \frac{4r}{\sqrt{AB}}\Bigl[rBA'\Bar{\psi}' + 2r\Bar{\psi}(e^2\Bar{A}^2+2\kappa B \Bar{A}'^2)+ A(-2m^2r\Bar{\psi} + rB'\Bar{\psi}'+ 2B(2\Bar{\psi}'+r\Bar{\psi}''))\Bigr]\,,\\
\label{eqAbar}
\mathcal{E}_{\Bar{A}} &\equiv  \frac{2r}{A\sqrt{AB}}\Bigl[-rB(1+4\kappa \Bar{\psi}^2)A' \Bar{A}' + A(rB'\Bar{A}' + 16r\kappa B \Bar{\psi} \Bar{\psi}'\Bar{A}'+
    + 2B(2\Bar{A}'+r\Bar{A}'') \nonumber\\
&\qquad + \Bar{\psi}^2(-4e^2r\Bar{A} + 4r\kappa B' \Bar{A}' + 8\kappa B (2\Bar{A}' + r\Bar{A}'')))\Bigr]\,.
\end{align}
where $'$ denotes derivative with respect to $r$. It will turn out to be useful to introduce the following functions
\begin{align}
Q(r) &= 4r^2\Bar{A}'\left(1+4\kappa \Bar{\psi}^2\right)\sqrt{\dfrac{B}{A}}\,,\qquad 
C(r) = \dfrac{e^2r^2\Bar{\psi}^2\Bar{A}^2}{\sqrt{AB}}\,,\\
J(r) &= 8r^2\sqrt{AB}\Bar{\psi}'\,,\qquad 
S(r) = \dfrac{8r^2\Bar{\psi}}{\sqrt{AB}}\left[-m^2A + e^2\Bar{A}^2+2\kappa B \Bar{A}'^2\right].
\end{align}
These functions will appear often while getting the perturbation equations and they are useful for rewriting $\mathcal{E}_{\Bar{\psi}}$ and $\mathcal{E}_{\Bar{A}}$ as follows
\begin{align}
\mathcal{E}_{\Bar{A}} &\equiv  Q'(r) - \dfrac{8}{\Bar{A}}C(r)\,,\quad 
\mathcal{E}_{\Bar{\psi}} \equiv  J'(r) + S(r).
\end{align}

\section{Existence of hairy warm holes}
\label{existence}
The gauge for the Maxwell potential is chosen such that $\bar{A}(r=r_H)=0$ where $r_H$ is the horizon radius and $\bar{A}(r=\infty)=\mu$. Under these conditions, and assuming the solution is asymptotically flat $A(r=\infty)=B(r=\infty)=1$, the scalar field eq.(\ref{eqPsi}) simplifies at infinity to
\begin{align}
\Bar{\psi}''+\frac{2}{r}\Bar{\psi}'+(e^2\mu^2-m^2) \Bar{\psi} = 0\,. 
\end{align}
This equation indicates that a localized scalar field, and consequently a configuration with finite energy, exists if and only if $e^2\mu^2\leq m^2$. When this inequality is saturated, the solution corresponds to a maximal warm hole \cite{Dias:2021vve}, which represents an extremal black hole with non-zero temperature.

\section{Background solution}

Following \cite{Dias:2021vve}, we compactify the space by introducing the new coordinate, $z=r_H/r$ which spans the range $[0,1]$ for the exterior solution. Additionally, we redefine the variables as follows:
\begin{align}
\label{relabel}
B(r) &= (1-z) q_1(z)\,,\quad A(r) = B(r) q_4(z)\,,\quad 
\bar{A} = (1-z) q_2(z)\,, \quad \bar{\psi} = q_3(z).
\end{align}
We also introduce the dimensionless parameter $y_H \equiv m r_H$. The resulting equations for the fields are expressed in terms of the new coordinate $z$ where $\dot{}$ denotes differentiation with respect to $z$

\begin{align}
\label{eq:q2}
& 2 (1-z) z^4 q_1 q_4 \left(1+4 \kappa  q_3^2\right)\ddot q_2-z^4 q_1 \Bigl[(1-z) \left(1+4 \kappa q_3^2\right) \dot q_4+4 q_4 \left(1+4 \kappa  q_3^2 + 4 \kappa  (z-1) q_3 \dot q_3\right)\Bigr]\dot q_2\nonumber\\
&\qquad +\Bigl[z^4 q_1 \left(1+4 \kappa  q_3^2\right) \dot q_4 - 4 q_3 q_4 \left(\frac{e^2}{m^2}y_H^2 q_3+4 \kappa z^4 q_1 \dot q_3\right)\Bigr]q_2=0\,,\\
\label{eq:q3}
&2 (1-z) z^4 q_1^2 q_4 \ddot q_3+z^4 q_1\Bigl[ q_1 \left((1-z) \dot q_4-2 q_4\right)+2 (1-z) q_4 \dot q_1\Bigr]\dot q_3+2\Bigl[\frac{e^2}{m^2}y_H^2(1-z)  q_2^2\nonumber\\
&\qquad +q_1 \left(2 \kappa  z^4 \left(q_2-(1-z) \dot q_2\right)^2-y_H^2 q_4\right)
\Bigr] q_3=0\,,\\
\label{eq:q4}
& z^3q_1^2(\dot{q}_4+4zq_4\dot{q}_3^2 )+4\frac{e^2}{m^2}y_H^2q_2^2q_3^2=0\,,\\
\label{eq:q1}
&2 q_4 \Bigl(z^2 \left(q_1-(1-z) z \dot q_1-1\right)+2 y_H^2 q_3^2\Bigr)+2 z^4 q_2 \Bigl(1+4 \kappa  q_3^2\Bigr)\left(q_2-2 (1-z) \dot q_2\right) \nonumber\\
&\qquad-(1-z) z^3 \Bigl(q_1 \dot q_4-2 (1-z) z \left(1+4 \kappa  q_3^2\right)\dot q_2^2 \Bigr)=0\,.
\end{align}
These equations depend on the ratio $e/m$ rather than the individual values of $e$ and $m$, meaning only this ratio affects the background solution. This system of equations could be solved provided 6 initial conditions but because only 4 are known, we will consider additional boundary conditions. At $z=0$, corresponding to $r\rightarrow \infty$, the fields satisfy the following boundary conditions:
\begin{align}
A=B=1\,,\quad \bar{\psi}=0\,,\quad \bar{A}=\mu\,,
\end{align}
or equivalently
\begin{align}
\label{bound:infy}
q_1(0) = q_4(0) = 1\,,\quad q_2(0) = \mu\,,\quad q_3(0)=0\,.
\end{align}
Regularity at the horizon, $z=1$, imposes additional conditions
\begin{align}
\label{bound:hori1}
&q_1 q_4- q_4+2y_+^2q_3^2q_4 + q_2^2(1+4\kappa q_3^2)=0\,,\\
\label{bound:hori2}
&q_1^2(1+4\kappa q_3^2)q_4^2\dot{q_2} - q_2q_3^2\Bigl(-4\kappa^2q_2^4(1+4\kappa q_3^2) - q_2^2\Bigl(\frac{e^2}{m^2} y_H^2 - 4y_H^2\kappa + 8\kappa^2q_1 + 4y_H^2\Bigl(\frac{e^2}{m^2}-4\kappa\Bigr)\kappa q_3^2\Bigr)q_4 \nonumber\\
&\qquad - y_H^2\Bigl(\Bigl(\frac{e^2}{m^2}-4\kappa\Bigr)q_1 + y_H^2(1+4\kappa q_3^2)\Bigr)q_4^2\Bigr)=0\,,\\
\label{bound:hori3}
&q_1 q_4 \dot{q_3} - q_3(2\kappa q_2^2-y_H^2q_4)=0\,,\\
\label{bound:hori4}
&q_1q_4\dot{q_4} +4 q_3^2\Bigl(4\kappa^2 q_2^4 + y_H^2\Bigl(\frac{e^2}{m^2}-4\kappa\Bigr)q_2^2q_4 + y_H^4q_4^2\Bigr)=0
\end{align}
All $q$-functions are evaluated at $z=1$. For any set of parameters $\{e/m, \kappa, y_H, \mu \}$, equations (\ref{eq:q2}, \ref{eq:q3}, \ref{eq:q4}, \ref{eq:q1}) are solved numerically as a boundary value problem (BVP) using the conditions in (\ref{bound:infy}) as asymptotic conditions and (\ref{bound:hori1}, \ref{bound:hori2}, \ref{bound:hori3}, \ref{bound:hori4}) as conditions at the horizon. Once solved, the original field functions $A(r)$, $B(r)$, $\bar{A}(r)$ and $\bar{\psi}(r)$ are recovered through the relations (\ref{relabel}). From these expressions, we can obtain the mass and charge of the black hole as
\begin{align}
\label{eq:MQ}
 M=\lim_{r\rightarrow \infty}   \frac{r^2 B'(r)}{2}\,,~~ Q_e=\lim_{r\rightarrow \infty}   r^2 \bar{A}'(r)\,.
\end{align}

The interval $0\leq z \leq 1$ is discretized using Chebyshev-Gauss-Lobatto collocation points
\begin{align}
\label{eq:secondkind}
    z_j =\frac{1}{2} \Bigl(1+\cos\Bigl(\frac{j}{N}\pi\Bigr)\Bigr)\,,\quad j\in \{0,1,\cdots,N\}
\end{align}
and the equations are linearized around an initial approximation $q_i^{(0)}$, with the following expansion
\begin{align}
    q_i(z) = q_i^{(0)}(z)+\delta q_i(z)\,.
\end{align}
Assuming an initial guess for $q_i^{(0)}$, the equations for $\delta q_i(z)$ are linear, reducing the problem to solving a matrix equation. Once $\delta q_i$ is obtained, the functions are updated as $q_i^{(1)}(z) = q_i^{(0)}(z)+\delta q_i(z)$. The new approximation $q_i^{(1)}$ serves as the seed for the next iteration
\begin{align}
    q_i(z) = q_i^{(1)}(z)+\delta q_i(z)\,.
\end{align}
This process is repeated until convergence is achieved, defined as $|\delta q_i(z)|<10^{-10}$. The primary challenge lies in selecting an appropriate initial guess. For most cases, the solution tends to converge to the Reissner-Nordström solution with a trivial scalar field with the following analytical form:
\begin{align}
\label{RN_Sol}
q_1(z) = 1-\mu^2z\,,\quad q_2(z) = \mu\,,  \quad
q_3(z) = 0\,, \quad q_4(z) = 1\,,
\end{align}
which is a valid solution that depends on the parameters $\mu$ and $r_H$, although uninteresting for our case. As expected, this solution respects the extremal bound $Q_e \leq M$ for any value of its parameters. Notice that RN black hole always exists for $Q_e\leq M$ but as shown in \cite{Dias:2021vve} it becomes unstable and gives rise to a hairy BH when $\alpha>1/8+y_H^2(1-\frac{e^2}{m^2})/2$ and $\mu$ surpasses a critical value.

To find non-trivial solutions, we used the \texttt{COLNEWSC} package \cite{colnew}. By using, as an initial guess, the Reissner-Nordström solution combined with a small constant for the scalar field, we successfully obtained non-trivial solutions. These non-trivial solutions were subsequently employed as seeds in our own code to verify the robustness of the procedure.

Once a solution was obtained, we used it as the initial guess to explore the parameter space $\{e/m, \kappa, y_H , \mu\}$ and construct solutions in all the parameter space.

\begin{figure}
    \centering
    \includegraphics[width=0.45\linewidth]{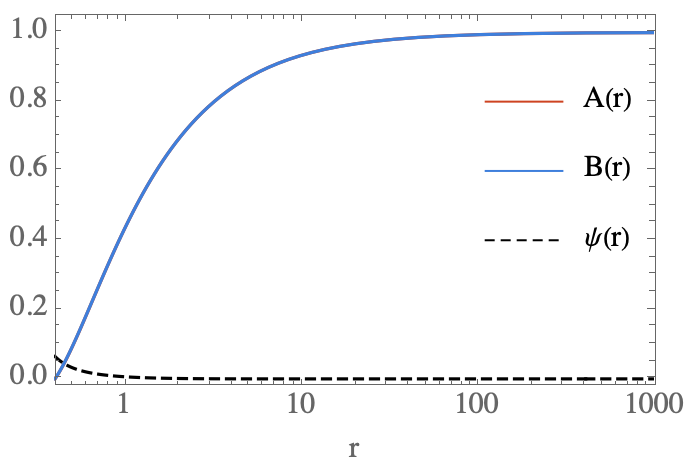}
    \includegraphics[width=0.45\linewidth]{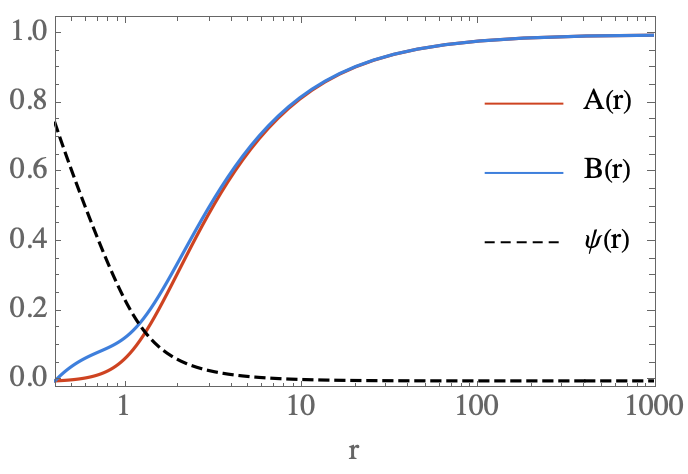}
    \caption{
        Potentials $A$ and $B$, along with the scalar field $\psi$, plotted as functions of the radial coordinate for a fixed horizon radius $r_H=0.4$, $\kappa=1$ and $e/m=1$. 
        In the first plot, $\mu=0.82$, which gives $M=0.336$ and $Q_e=0.33$ according to eqs. (\ref{eq:MQ}). 
        In the second plot, $\mu=1$, corresponding to $M=0.977$ and $Q_e=0.999$. The presence of the scalar field clearly generates a difference between the 2 metric potentials $A(r)$ and $B(r)$.
    }
    \label{fig:background}
\end{figure}

Figure (\ref{fig:background}) presents two distinct plots comparing black holes that either respect or violate the usual extremal bound $Q_e=M$. In the first plot, corresponding to $Q_e-M=-0.006$, the difference between the potentials $A$ and $B$ is negligible, except very close to the horizon, while the scalar field remains very small. In contrast, the second plot shows a case where the usual extremal bound is violated, with $Q_e-M=0.023$. Here, the scalar field is larger, and the difference between the two potentials becomes clearly apparent.

These background solutions serve as the basis for analyzing the stability of the black holes.

\section{Perturbation formalism}

In this section, we summarize the formalism for perturbations around a static spherically symmetric spacetime based on the work initiated by \cite{Regge:1957td,Zerilli:1970se}. Our approach follows very closely \cite{Gannouji:2021oqz}. We consider a background defined by eqs.(\ref{line},\ref{back_fields}) to which we add a perturbation
\begin{align}
g_{\mu\nu} &=\bar g_{\mu\nu}+h_{\mu\nu} \,,\quad
A_\mu = \bar A_\mu +\delta A_\mu \,,\quad
\psi = \bar \psi +\delta \psi \,,\quad
\psi^\dag = \bar \psi^\dag +\delta \psi^\dag \,.
\end{align}
Each field can be decomposed into scalar and vector perturbations which decouple at the linear regime and therefore can be studied separately. 

In summary, we have four vector perturbations:

\begin{align}
    h_{0i} &= \sum_{l, m}h^{(0)}_{\ell m}(t, r)E_{ij}\partial^j Y_\ell^m(\theta, \phi),\quad
    h_{1i} = \sum_{\ell, m}h^{(1)}_{\ell m}(t, r)E_{ij}\partial^j Y_\ell^m(\theta, \phi),\\
    h_{ij} &= \dfrac{1}{2}\sum_{\ell, m}h^{(2)}_{\ell m}(t, r)\left[E_{i}^k\nabla_{kj} Y_\ell^m(\theta, \phi) + E_{j}^k\nabla_{ki} Y_\ell^m(\theta, \phi) \right],\quad
    \delta A_i = \sum_{\ell, m}A^{(v)}_{\ell m}(t, r)E_{ij}\nabla^{j} Y_\ell^m(\theta, \phi),
\end{align}
where the indices $i,j$ take values $\{2,3\}$, $Y_\ell^m$ are the spherical harmonics, $E_{i j}=\sin\theta \epsilon_{i j}$ and $\epsilon_{i j}$ is the Levi-Civita symbol with
$\epsilon_{23}$ = 1. Due to the spherical symmetry of the background spacetime, we can, without loss of generality, assume $m=0$.

\noindent We have also 12 scalar perturbations, namely
\begin{align}
    h_{00} &= A(r)\sum_{\ell, m}H^{(0)}_{\ell m}(t, r)Y_\ell^m(\theta, \phi),\quad
    h_{01} = \sum_{\ell, m}H^{(1)}_{\ell m}(t, r)Y_\ell^m(\theta, \phi),\\
    h_{11} &= \dfrac{1}{B(r)}\sum_{\ell, m}H^{(2)}_{\ell m}(t, r)Y_\ell^m(\theta, \phi),\quad
    h_{0i} = \sum_{\ell, m}\beta_{\ell m}(t, r)\partial_iY_\ell^m(\theta, \phi),\\
    h_{1i} &= \sum_{\ell, m}\alpha_{\ell m}(t, r)\partial_iY_\ell^m(\theta, \phi),\quad
    h_{ij} = \sum_{\ell, m}\left[K_{\ell m}(t, r)\gamma_{ij}Y_\ell^m(\theta, \phi) + G_{\ell m}(t, r)\nabla_{ij}Y_\ell ^m(\theta, \phi) \right],
\end{align}
for the metric sector and for the other fields we have
\begin{align}
    \delta A_0 &= \sum_{\ell, m}A^{(0)}_{\ell m}(t, r)Y_\ell^m(\theta, \phi),\quad 
    \delta A_1 = \sum_{\ell, m}A^{(1)}_{\ell m}(t, r)Y_\ell^m(\theta, \phi),\quad
    \delta A_i = \sum_{\ell, m}A^{(2)}_{\ell m}(t, r)\partial_{i} Y_\ell^m(\theta, \phi),\nonumber\\
    \delta \psi &= \sum_{\ell, m}\delta \psi_{\ell m}(t, r)Y_\ell^m(\theta, \phi),\quad
    \delta \psi^\dagger = \sum_{\ell, m}\delta \psi^\dagger_{\ell m}(t, r)Y_\ell^m(\theta, \phi).\label{s11}
\end{align}

Of course, we can use the gauge freedom to eliminate some of the variables. This choice depends on the mode $\ell$. To simplify the discussion, we will summarize only the case $\ell\geq 2$. For the discussion of the gauge in all cases see \cite{Gannouji:2021oqz}.

For the vector sector, and when $\ell\geq 2$, the only full gauge fixing is the Regge-Wheeler gauge, namely $h_{\ell m}^{(2)}=0$. For the scalar perturbations we have different possible gauges. We chose $G_{\ell m} = K_{\ell m} = A_{\ell m}^{(2)} = \delta \psi_{\ell m}^{\dagger}=0$. Notice that we can't use this gauge to study the case $\bar{\psi}=0$ which corresponds to Reissner-Nordström black hole. Indeed, if $\bar{\psi}=0$, the perturbation is gauge invariant according to the Stewart-Walker lemma \cite{Stewart:1974uz} and therefore the gauge choice $\delta \psi_{\ell m}^{\dagger}=0$ is not allowed.

\section{Vector Perturbations}

In the vector sector of perturbations, the mode $\ell$ can be larger than $2$ but we can also analyse the dipole perturbation $\ell=1$. Focusing, first on the general case $\ell\geq 2$ and expanding the action to second order and integrating over ($\theta, \phi$) we get
\begin{align}
\label{accion_vec}
S_{\mathrm{vec}}^{(2)} = \dfrac{2\ell+1}{4\pi}\int {\mathrm  d}t{\mathrm  d}r \mathcal{L}_{\mathrm{vec}}^{(2)},
\end{align}
where the Lagrangian density $\mathcal{L}_\mathrm{vec}^{(2)}$ after some integrations by parts is
\begin{align}
\label{L_vec1}
&\mathcal{L}_{\mathrm{vec}}^{(2)} = a_1 h_0^2 + a_2 h_1^2 + a_3 \Bigl[ \dot{h}_1^2 + h_0'^{2} + \frac{4}{r}h_0\Dot{h}_1 - 2h'_0\dot{h}_1 + 2a_4 \left(h'_0 - \dot{h}_1\right)A_v\Bigr]+ a_5 h_0 A_v + a_6 A_v^2\nonumber\\ 
&\qquad + a_7 \Dot{A}_v^2 + a_8 A_v'^2
\end{align}
For simplicity of the notation, we have renamed the functions such as $h_{lm}^{(0)}\rightarrow h_0$, $h_{lm}^{(1)}\rightarrow h_1$ and $A_{lm}^{(v)}\rightarrow A_v$. The coefficients are given by
\begin{align}
    a_1&=\dfrac{\lambda}{2r^2}\left[\left(2r\sqrt{\dfrac{B}{A}}\right)' + \dfrac{\lambda-2}{\sqrt{AB}}\right],\quad
    a_2=-\dfrac{\sqrt{AB}}{2r^2}\lambda (\lambda-2),\quad
    a_3=\dfrac{\lambda}{2}\sqrt{\dfrac{B}{A}},\quad
    a_4=\sqrt{\dfrac{A}{B}}\dfrac{Q}{r^2},,\nonumber\\
    a_5 &=-\frac{2\lambda Q}{r^3}\,,\quad a_6=-4e^2\lambda \sqrt{\dfrac{A}{B}} \Bar{\psi}^2 - \dfrac{A\lambda^2 Q}{2B\Bar{A}'r^4},\quad
    a_7=\dfrac{\lambda Q}{2B\Bar{A}'r^2},\quad
    a_8=-\dfrac{\lambda Q A}{2\Bar{A}'r^2},
\end{align}
with $\lambda = \ell(\ell+1)$. The Lagrangian (\ref{L_vec1}) can be written as
\begin{align}
\label{L_vec2}
\mathcal{L}_{\mathrm{vec}}^{(2)} &= b_1 h_0^2 + a_2 h_1^2 + a_3 \Bigl[ \Dot{h}_1 - h_0' + \dfrac{2}{r}h_0 - a_4A_v\Bigr]^2 + b_2 h_0 A_v + \left(a_6 - a_3 a_4^2\right) A_v^2 + a_7 \Dot{A}_v^2 + a_8 A_v'^2\,,
\end{align}
where $b_1 = a_1 - (2ra_3)'/r^2$ and $b_2 = a_5 + 4 a_3 a_4/r = \lambda \mathcal{E}_{\Bar{A}}/r^2$, so it vanishes on-shell. We introduce an auxiliary field $q(t, r)$ and define an equivalent Lagrangian
\begin{align}
\label{L_vec3}
\mathcal{L}_{\mathrm{vec}}^{(2)} &= b_1 h_0^2 + a_2 h_1^2 + a_3 \Bigl[2q\left(\Dot{h}_1 - h_0' + \dfrac{2}{r}h_0 - a_4A_v\right) - q^2 \Bigr] + \left(a_6 - a_3 a_4^2\right) A_v^2 + a_7 \Dot{A}_v^2 + a_8 A_v'^2\,.
\end{align}
Taking the variation of this action with respect to $h_0$ and $h_1$, we obtain
\begin{align*}
    h_0 = -\dfrac{(a_3 q r^2)'}{b_1 r^2}\,,\quad  h_1 = \dfrac{a_3}{a_2}\Dot{q}.
\end{align*}
Substituting these expressions in the Lagrangian we obtain
\begin{align}
\label{L_vec4}
\mathcal{L}_{\mathrm{vec}}^{(2)} =& \alpha_1 \Dot{q}^2 + \beta_1 q'^2 + \gamma_1 q^2 + \alpha_2 \Dot{A}_v^2 + \beta_2 A_v'^2 + \gamma_2 A_v^2 + \sigma A_vq.
\end{align}
The number of functions that appear now reflect the number of degrees of freedom that propagate in the vector perturbation sector. One degree comes from the gravitational field while the other comes from the Maxwell field. The coefficients of the above Lagrangian are the following
\begin{align*}
\alpha_1 &=\frac{r^2 \sqrt{B}}{2A^{3/2}}\frac{\lambda}{\lambda-2}\,,\quad 
\alpha_2 =\dfrac{\lambda Q}{2B\Bar{A}'r^2}\,,\quad 
\sigma=-\frac{Q\lambda}{r^2}\,,\quad
\beta_1=-AB\alpha_1\,,\quad 
\beta_2=-AB\alpha_2\,,\\
\gamma_2 &=a_6 -a_3a_4^2\,,\quad
\gamma_1 =\dfrac{a_3}{b_1^2r^4}\Bigl[-r^2b_1'(a_3r^2)' + b_1 \left[r^2 (a_3r^2)'' - 4r(a_3r^2)'\right] - b_1^2 r^4\Bigr]
\end{align*}
In order to write the perturbation in a canonical form, we introduce the tortoise coordinate $dr = \sqrt{AB} dr_*$ and the new variables
\begin{align}
\label{rescale_v}
q(t, r) &= \sqrt{\dfrac{A}{2Br^2}}V_g(t, r)\,,\quad
A_v(t, r) = \dfrac{V_e(t, r)}{\sqrt{8(\lambda-2)(1+4\kappa \Bar{\psi}^2)}}.
\end{align}
The action becomes
\begin{align}
\label{accion_vec_final}
S_{\mathrm{vec}}^{(2)} &= \dfrac{\ell(\ell+1)(2\ell+1)}{16\pi(\ell+2)(\ell-1)}\int {\mathrm  d}t {\mathrm  d} r_* \Bigl[\left(\dfrac{\partial \Psi_i}{\partial t}\right)^2 - \left(\dfrac{\partial \Psi_i}{\partial r_*}\right)^2- V_{ij}\Psi_i\Psi_j\Bigr],
\end{align}
where $\vec{\Psi}= (V_g, V_e)^T$ and $V_{ij}$ are the components of a $2 \times 2$ symmetric matrix 
\begin{align}
V_{11}&=(\lambda-2)\dfrac{A}{r^2} - \partial_{r_*}S_1 + S_1^2\,,\quad
V_{12} =V_{21} = \sqrt{(\lambda-2)\dfrac{A}{r^2}G(r)}\,,\nonumber\\
V_{22}&=\dfrac{2e^2A \Bar{\psi}^2}{(1+4\kappa \Bar{\psi}^2)} + \dfrac{\lambda A}{r^2} + G(r) - \partial_{r_*}S_2 + S_2^2
\label{pot:vl2}
\end{align}
where
\begin{align}
S_1(r) =\dfrac{\sqrt{AB}}{r}\,,\quad
S_2(r)=-\dfrac{4\kappa \sqrt{AB}\Bar{\psi}\Bar{\psi}'}{(1+4\kappa \Bar{\psi}^2)}\,,\quad
G(r) =\dfrac{AQ^2}{4r^4(1+4\kappa \Bar{\psi}^2)}\,.
\end{align}

The variation of the action (\ref{accion_vec_final}) yields a system of two coupled linear wave-like differential equations
\begin{align}
&\frac{\partial^2 \Psi_1}{\partial t^2}-\frac{\partial^2 \Psi_1}{\partial r_*^2}+V_{11}\Psi_1+V_{12}\Psi_2 = 0 \,,\\
&\frac{\partial^2 \Psi_2}{\partial t^2}-\frac{\partial^2 \Psi_2}{\partial r_*^2}+V_{22}\Psi_2+V_{12}\Psi_1 = 0\,,
\end{align}
with $V_{12}$ acts as the coupling term between the two perturbations.

Following \cite{Gannouji:2021oqz} we can perform an S-deformation from which we can conclude that the vector perturbation is stable if $G(r)$ is positive which is trivially satisfied.

We will now study the dipole perturbation, $\ell=1$. In that case, the perturbation $h^{(2)}_{lm}$ is identically zero and therefore we do not use the Regge-Wheeler gauge. We will use another gauge later. For the moment, we follow the same procedure and rescaling the Maxwell potential perturbation as follows
\begin{align}
\label{rescale_l1}
A_v(t, r) = \dfrac{V_e(t, r)}{\sqrt{4(1+4\kappa \Bar{\psi}^2)}},
\end{align}
we obtain
\begin{align}
\label{L_vec_l1_2}
\mathcal{L}_{\mathrm{vec,}\hspace{1mm}l=1}^{(2)} = & b_1 \left[ \Dot{h}_1 - h_0' + \dfrac{2}{r}h_0 - b_2V_e\right]^2  + b_3 V_e^2+ b_4 \Dot{V}_e^2 + b_5 V_e'^2,
\end{align}
where the coefficients are
\begin{align}
b_1&=\sqrt{\dfrac{B}{A}}\,, \quad    b_2=2\sqrt{1+4\kappa \Bar{\psi}^2}\Bar{A}',\quad
b_4=\dfrac{1}{\sqrt{AB}}\,,\quad b_5 =-\sqrt{AB}\,,\nonumber\\
b_3&=\dfrac{1}{r^2\sqrt{AB}(1+4\kappa \Bar{\psi}^2)} \Bigl[4r^2(1+4\kappa \Bar{\psi}^2)(e^2\kappa\Bar{A}^2\Bar{\psi}^2-B(1+2\kappa \Bar{\psi}^2(4-\kappa + 8\kappa\Bar{\psi}^2))\Bar{A}'^2)\nonumber\\
&\quad + 2A(-((1+4\kappa \Bar{\psi}^2)(1+(4\kappa+r^2(e^2+2m^2\kappa))\Bar{\psi}^2)) + 4r\kappa B\Bar{\psi}(1+4\kappa \Bar{\psi}^2)\Bar{\psi}' - 2r^2 \kappa B \Bar{\psi}'^2)\Bigr].
\end{align}
Variation of the action with respect to $h_0$ and $h_1$ gives  
\begin{align}
\label{K_eqs}
\Dot{\mathcal{K}} = 0\,,\quad  (r^2\mathcal{K})' = 0,
\end{align}
where we have introduced 
\begin{align}
\label{K_def}
\mathcal{K} = b_1 \left(\Dot{h}_1 - h_0' + \dfrac{2}{r}h_0 - b_2V_e\right).
\end{align}
The solution of eq.\eqref{K_eqs} is $\mathcal{K} = \mathcal{J}/r^2$, where $\mathcal{J}$ is an integration constant. 

Remember that we didn't fix the gauge yet. Unfortunately, there is no complete gauge fixing. We use a partial gauge freedom to take $h_1=0$, but this induces a residual gauge of the form $h_0 \rightarrow h_0 - r^2f(t)$ with $f(t)$ an arbitrary real function. Taking \eqref{K_def} and the solution of eq.\eqref{K_eqs} and considering $h_1=0$, we obtain
\begin{align}
\label{K_join}
b_1 \left(- h_0' + \dfrac{2}{r}h_0 - b_2V_e\right) = \dfrac{\mathcal{J}}{r^2}.
\end{align}
Solving this equation for $h_0$ we get
\begin{align}
h_0(t, r) = -\mathcal{J}r^2 \int \dfrac{1}{b_1r^4}{{\mathrm  d}}r - r^2\int\dfrac{b_2}{r^2}V_e {{\mathrm  d}}r + r^2 F(t),
\end{align}
where $F(t)$ is an integration constant, that can be eliminated using the residual gauge, to obtain
\begin{align}
h_0(t, r) = -\mathcal{J}r^2 \int \sqrt{\dfrac{A}{B}}\dfrac{{\mathrm  d}r}{r^4} - 2r^2\int\dfrac{\sqrt{1+4\kappa \Bar{\psi}^2}\Bar{A}'}{r^2}V_e {\mathrm  d}r
\end{align}
In conclusion, $h_0$ can be obtained once $A_v$ is known. The action for $\ell=1$ and $h_1=0$ gives
\begin{align}
S_{\mathrm{vec},\:\ell=1} \propto &\int {\mathrm  d}t{\mathrm  d}r_* \Bigl[\left(\dfrac{\partial V_e}{\partial t}\right)^2 - \left(\dfrac{\partial V_e}{\partial r_*}\right)^2 - V_{22} V_e^2\nonumber- \sqrt{AB} b_1 \left(-h_0' + \dfrac{2}{r}h_0 -b_2V_e\right)^2\Bigr].
\end{align}
The variation of this action with respect to $V_e$ and considering \eqref{K_join} we obtain
\begin{align}
\label{eq:vecl1}
-\dfrac{\partial^2 V_e}{\partial t^2} + \dfrac{\partial^2 V_e}{\partial r_*^2} - V_{22, \:\ell=1}V_e - \dfrac{\sqrt{AB}b_2\mathcal{J}}{r^2} = 0.
\end{align}
We have a wave-like equation for $V_e$ with a non-homogeneous term. The homogeneous solution will describe the propagation of the electromagnetic perturbation in our spacetime at the speed of light with a potential $V_{22,\:\ell=1}$ that is the same as in (\ref{pot:vl2}), with $\ell=1$. Similarly to the previous case, the S-deformation proves that the perturbation is stable. 

In conclusion, independently of the background solution, the black hole is stable under all vector perturbations, $\ell\geq 1$.

\section{Scalar perturbations}

In this section, we study the scalar perturbations. As before we redefine our variables such that $H_{\ell m}^{(0)}\rightarrow H_0$, $H_{\ell m}^{(1)}\rightarrow H_1$, $H_{\ell m}^{(2)}\rightarrow H_2$, $\alpha_{\ell m}\rightarrow \alpha$, $\beta_{\ell m}\rightarrow \beta$, $A_{\ell m}^{(0)}\rightarrow A_0$, $A_{\ell m}^{(1)}\rightarrow A_1$ and $\delta \psi_{\ell m}\rightarrow \delta \psi$. Expanding the action at the second order, we arrive at
\begin{align}
\label{accion_sca}
S_{\mathrm{sca}}^{(2)} = \dfrac{2\ell+1}{\pi}\int {\mathrm  d}t{\mathrm  d}r \mathcal{L}_{\mathrm{sca}}^{(2)},
\end{align}
with
\begin{align}
\label{L_sca1}
\mathcal{L}_{\mathrm{sca}}^{(2)} &= a_1 H_0^2 + a_2H_0\alpha + a_3H_0\delta\psi + a_4H_0A_0 + a_5H_0\alpha'+ a_6H_0\delta\psi' + a_7H_0(A_0' - \Dot{A}_1) + b_1 H_1^2 + b_2H_1\beta\nonumber\\
&+ b_3H_1\beta' + b_4H_1\Dot{\alpha} + b_5H_1\Dot{\delta\psi} + b_6H_1A_1 + c_1 H_2^2+ H_2\Bigl[c_2H_0 + c_3A_0 + c_4\alpha + c_5\delta\psi + c_6(A_0' - \Dot{A}_1)\nonumber\\
&+ c_7 H_0' + c_8\Dot{H}_1 + c_9\delta\psi' + c_{10}\Dot{\beta}\Bigr] + d_1\alpha^2 + d_2A_0\alpha+ d_3\alpha\delta\psi + d_4\beta^2 + d_5A_1\beta + d_6\beta'^2 + d_7 \beta'\Dot{\alpha} + d_8\Dot{\alpha}^2\nonumber\\
&+ d_9\beta\Dot{\alpha} + e_1\delta\psi^2 + e_2A_0\delta\psi + e_3\delta\psi(A_0' - \Dot{A}_1) + e_4\delta\psi'^2+ e_5 \Dot{\delta\psi}^2 + f_1A_0^2 + f_2A_1^2 + f_3(A_0' - \Dot{A}_1)^2,
\end{align}

\noindent
where $a_i$, $b_i$, $c_i$, $d_i$, $e_i$ and $f_i$ are functions of $r$ and their expressions are given in Appendix \ref{appendix:coeff}. The objective is to reduce the number of functions in our Lagrangian from eight to four, so they reflect the number of degrees of freedom remaining in the theory. Variation with respect to $A_0$ and $A_1$ gives us, respectively
\begin{align}
&\partial_r\left[a_7H_0 + c_6H_2 + e_3\delta\psi + 2f_3(A_0' - \Dot{A}_1)\right] - a_4H_0 - c_3H_2 - d_2\alpha - e_2\delta\psi - 2f_1A_0 = 0, \\
&\partial_t\left[a_7H_0 + c_6H_2 + e_3\delta\psi + 2f_3(A_0' - \Dot{A}_1)\right] + b_6H_1+ d_5\beta + 2f_2A_1 = 0.
\end{align}
Following \cite{Gannouji:2021oqz}, it turns out to be useful to introduce a new variable
\begin{align}
\label{elec_Se}
S_e(t, r) = a_7H_0 + c_6H_2 + e_3\delta\psi + 2f_3(A_0' - \Dot{A}_1),
\end{align}
that allows us to get algebraic expressions for $A_0$ and $A_1$
\begin{align}
\label{elec_Se_1}
&A_0 = \dfrac{1}{2f_1} \left(S_e' - a_4H_0 - c_3H_2 - d_2\alpha - e_2\delta\psi \right)\,,\\
\label{elec_Se_111}
&A_1 = -\dfrac{1}{2f_2}\left(\Dot{S}_e + b_6H_1 + d_5\beta\right).
\end{align}
Replacing \eqref{elec_Se}, \eqref{elec_Se_1} and \eqref{elec_Se_111} in our Lagrangian \eqref{L_sca1} we get 
\begin{align}
\label{L_sca2}
\mathcal{L}_{\mathrm{sca}}^{(2)} &= a_1 H_0^2 + a_2H_0\alpha + a_3H_0\delta\psi + a_4H_0\alpha' + a_5H_0\delta\psi'+ b_1 H_1^2 + b_2H_1\beta + b_3H_1\beta' + b_4H_1\Dot{\alpha} + b_5H_1\Dot{\delta\psi}\nonumber\\
&+ c_1 H_2^2 + H_2\Bigl[c_2H_0 + c_3\alpha + c_4\delta\psi + c_5 H_0' + c_6\Dot{H}_1+ c_7\delta\psi' + c_8\Dot{\beta}\Bigr] + d_1\alpha^2 + d_2\alpha\delta\psi + d_3\beta^2 + d_4\beta'^2\nonumber\\
&+ d_5\beta'\Dot{\alpha} + d_6\Dot{\alpha}^2 + d_7\beta\Dot{\alpha} + e_1\delta\psi^2 + e_2\delta\psi'^2 + e_3 \Dot{\delta\psi}^2+ e_4S_e^2 + e_5S_e'^2 + e_6\Dot{S}_e^2,
\end{align}
where the coefficients are now redefined and they are not very illuminating at this stage. Notice that function $S_e$ decouples from the rest and therefore we will omit it in the next steps. Also, there are no derivatives of $H_2$ in our Lagrangian, so we can vary the action with respect to it and get the following algebraic equation
\begin{equation}\label{eqH2}
    2c_1H_2 + c_2H_0 + c_3\alpha + c_4\delta\psi + c_5 H_0' + c_6\Dot{H}_1 + c_7\delta\psi'+ c_8\Dot{\beta}=0.
\end{equation}

\noindent
By introducing a new function

\begin{equation}\label{S1_intro}
    S_1(t, r) = H_1 + \dfrac{c_8}{c_6}\beta,
\end{equation}

\noindent
in equation (\ref{eqH2}) we get an algebraic expression for $H_2$

\begin{equation}\label{H2_out}
    H_2=\dfrac{1}{2c_1}\left[c_2H_0 + c_3\alpha + c_4\delta\psi + c_5 H_0' + c_6\Dot{S}_1 + c_7\delta\psi'\right].
\end{equation}

\noindent
By eliminating $H_1$ in the Lagrangian \eqref{L_sca2} using \eqref{S1_intro} and then eliminating $H_2$ using \eqref{H2_out} we get (after some integration by parts)
\begin{align}
\label{L_sca3}
\mathcal{L}_{\mathrm{sca}}^{(2)} &= a_1 H_0^2 + a_2H_0\alpha + a_3H_0\delta\psi + a_4\alpha H_0' + a_5 H_0 \Dot{S}_1+ a_6H_0'\Dot{S}_1 + a_7 H_0'^2 + a_8H_0'\delta\psi' + a_9H_0'\delta\psi + a_{10}H_0\delta\psi'\nonumber\\
&+ b_1 S_1^2 + b_2S_1\beta + b_3\alpha \Dot{S}_1 + b_4 \Dot{S}_1^2 + b_5\Dot{S}_1\delta\psi' + b_6 \Dot{S}_1\delta\psi+ b_7S_1 \Dot{\delta\psi} + c_1S_e^2 + c_2S_e'^2 + c_3\Dot{S}_e^2 + d_1\alpha^2 + d_2\beta^2\nonumber\\
&+ d_3\alpha\delta\psi + d_4\alpha\delta\psi' + d_5\beta\Dot{\delta\psi} + d_6\Bigl[\beta'^2 - 2\beta'\Dot{\alpha} + \Dot{\alpha}^2-2S_1\Dot\alpha - 2S_1\beta'\Bigr] + d_7\beta\Dot{\alpha} + e_1\delta\psi^2 + e_2\delta\psi'^2 + e_3 \Dot{\delta\psi}^2\,.
\end{align}
The coefficients are again redefined and do not correspond to the ones in the previous Lagrangian. Their particular values are not important at this stage. Notice that the term in brackets is a square of the derivatives of $\alpha$ and $\beta$ as follows
\begin{align}
\label{L_sca4}
\mathcal{L}_{\mathrm{sca}}^{(2)} &= a_1 H_0^2 + a_2H_0\alpha + a_3H_0\delta\psi + a_4\alpha H_0' + a_5 H_0 \Dot{S}_1+ a_6H_0'\Dot{S}_1 + a_7 H_0'^2  + a_8H_0'\delta\psi' + a_9H_0'\delta\psi + a_{10}H_0\delta\psi' \nonumber\\
&+ (b_1 - d_6) S_1^2 + (b_2 - d_7) S_1\beta + (b_3 + 4d_6)\alpha \Dot{S}_1+ b_4 \Dot{S}_1^2 + b_5\Dot{S}_1\delta\psi' + b_6 \Dot{S}_1\delta\psi + b_7S_1 \Dot{\delta\psi} + c_1S_e^2 + c_2S_e'^2\nonumber\\
& + c_3\Dot{S}_e^2 + d_1\alpha^2 + d_3\alpha\delta\psi + d_4\alpha\delta\psi' + \left(d_2 - \dfrac{d_7^2}{4d_6} - \dfrac{d_7'}{2}\right) \beta^2 + d_5\beta\Dot{\delta\psi} + d_6\left[\beta' - \Dot{\alpha} - S_1 - \dfrac{d_7}{2d_6}\beta\right]^2\nonumber\\
&+ e_1\delta\psi^2 + e_2\delta\psi'^2 + e_3 \Dot{\delta\psi}^2.
\end{align}
As in the vector perturbation sector, we introduce an auxiliary field $S_g$ and define the equivalent Lagrangian
\begin{align}
\label{L_sca5}
\mathcal{L}_{\mathrm{sca}}^{(2)} &= a_1 H_0^2 + a_2H_0\alpha + a_3H_0\delta\psi + a_4\alpha H_0' + a_5 H_0 \Dot{S}_1+ a_6H_0'\Dot{S}_1 + a_7 H_0'^2 + a_8H_0'\delta\psi' + a_9H_0'\delta\psi + a_{10}H_0\delta\psi' \nonumber\\
&+ (b_1 - d_6) S_1^2 + (b_2 - d_7) S_1\beta + (b_3 + 4d_6)\alpha \Dot{S}_1 + b_4 \Dot{S}_1^2+ b_5\Dot{S}_1\delta\psi' + b_6 \Dot{S}_1\delta\psi + b_7S_1 \Dot{\delta\psi} + c_1S_e^2 + c_2S_e'^2 \nonumber\\
&+c_3\Dot{S}_e^2+ d_1\alpha^2 + d_3\alpha\delta\psi + d_4\alpha\delta\psi' + \left(d_2 - \dfrac{d_7^2}{4d_6} - \dfrac{d_7'}{2}\right) \beta^2 \nonumber+ d_5\beta\Dot{\delta\psi} + d_6\Bigl[2S_g\left(\beta' - \Dot{\alpha} - S_1 - \dfrac{d_7}{2d_6}\beta\right)\nonumber\\
& - S_g^2\Bigr] + e_1\delta\psi^2 + e_2\delta\psi'^2 + e_3 \Dot{\delta\psi}^2.
\end{align}
Taking the variation with respect to $\alpha$ and $\beta$, we get algebraic expressions for both variables
\begin{align}
\alpha &= \dfrac{a_2H_0 + a_4H_0' + (b_4 + 4d_6)\Dot{S}_1 + d_3 \delta\psi + d_4 \delta\psi' + 2d_6\Dot{S}_g}{2d_1}\\
\beta &= \dfrac{(b_2 - d_7) S_1 + d_5\Dot{\delta\psi}-(d_7- 2d_6')S_g - 2 d_6 S_g'}{2\left(d_2 - \dfrac{d_7^2}{4d_6} - \dfrac{d_7'}{2}\right)}\,,
\end{align}
which after some integration by parts, reduce our Lagrangian to
\begin{align}
\label{L_sca6}
\mathcal{L}_{\mathrm{sca}}^{(2)} &= a_1 H_0^2 + a_2H_0\delta\psi + a_3H_0\delta\psi' + a_4H_0\Dot{S}_g + b_1 S_1^2 + b_2 S_1S_g +b_3 S_1 \Dot{\delta\psi} + b_4 S_1 S_g' + c_1 S_g^2 + c_2 S_g'^2\nonumber\\
&+ c_3 \Dot{S}_g^2 + c_4\delta\psi \Dot{S}_g + c_5\delta\psi' \Dot{S}_g + d_1S_e^2 + d_2S_e'^2 + d_3\Dot{S}_e^2 + e_1\delta\psi^2 + e_2\delta\psi'^2 + e_3 \Dot{\delta\psi}^2 + f_1 \left(H_0' - \dfrac{2}{A}\Dot{S}_1\right)^2\nonumber\\
&+ f_2 \left(H_0' - \dfrac{2}{A}\Dot{S}_1\right)\Dot{S}_g + f_3 H_0 \Dot{S}_1 + f_4 \left(H_0' - \dfrac{2}{A}\Dot{S}_1\right) \delta\psi'+ f_5 \left(H_0' - \dfrac{2}{A}\Dot{S}_1\right)\delta\psi,
\end{align}
Note again that we have redefined our coefficients, they do not correspond to the original ones. In order to use the same path for the variables $H_0$ and $S_1$, we first rewrite our Lagrangian in the following form
\begin{align}
\label{L_sca7}
\mathcal{L}_{\mathrm{sca}}^{(2)} &= \left(a_1 - \dfrac{A^2f_3^2}{16 f_1} - \dfrac{(Af_3)'}{4}\right) H_0^2 + \left(a_2 + \dfrac{Af_3f_5}{4f_1}\right)H_0\delta\psi+ \left(a_3 + \dfrac{Af_3f_4}{4f_1}\right)H_0\delta\psi' + \left(a_4 + \dfrac{Af_3f_2}{4f_1}\right)H_0\Dot{S}_g\nonumber\\
&+ b_1 S_1^2 + b_2 S_1S_g + b_3 S_1 \Dot{\delta\psi} + b_4 S_1 S_g' + c_1 S_g^2 + c_2 S_g'^2 + \left(c_3 - \dfrac{f_2^2}{4f_1}\right) \Dot{S}_g^2 + \left(c_4-\dfrac{f_2f_5}{2f_1}\right)\delta\psi \Dot{S}_g\nonumber\\
&+ \left(c_5-\dfrac{f_2f_4}{2f_1}\right)\delta\psi' \Dot{S}_g + d_1S_e^2 + d_2S_e'^2 + d_3\Dot{S}_e^2 + \left(e_1-\dfrac{f_5^2}{4f_1} + \dfrac{1}{4}\left(\dfrac{f_4f_5}{f_1}\right)'\right)\delta\psi^2  + \left(e_2-\dfrac{f_4^2}{4f_1}\right)\delta\psi'^2\nonumber\\
&+ e_3 \Dot{\delta\psi}^2 + f_1 \Bigl[H_0' - \dfrac{2}{A}\Dot{S}_1 + \dfrac{f_2}{2f_1}\Dot{S}_g + \dfrac{f_4}{2f_1}\delta\psi' + \dfrac{f_5}{2f_1}\delta\psi - \dfrac{Af_3}{4f_1}H_0\Bigr]^2.
\end{align}
We can finally introduce an other auxiliary field $S_h$ and the following equivalent Lagrangian
\begin{align}
\label{L_sca8}
\mathcal{L}_{\mathrm{sca}}^{(2)} &= \left(a_1 - \dfrac{A^2f_3^2}{16 f_1} - \dfrac{(Af_3)'}{4}\right) H_0^2 + \left(a_2 + \dfrac{Af_3f_5}{4f_1}\right)H_0\delta\psi+ \left(a_3 + \dfrac{Af_3f_4}{4f_1}\right)H_0\delta\psi' + \left(a_4 + \dfrac{Af_3f_2}{4f_1}\right)H_0\Dot{S}_g \nonumber\\
&+ b_1 S_1^2 + b_2 S_1S_g + b_3 S_1 \Dot{\delta\psi} + b_4 S_1 S_g' + c_1 S_g^2 + c_2 S_g'^2 + \left(c_3 - \dfrac{f_2^2}{4f_1}\right) \Dot{S}_g^2 + \left(c_4-\dfrac{f_2f_5}{2f_1}\right)\delta\psi \Dot{S}_g\nonumber\\
&+ \left(c_5-\dfrac{f_2f_4}{2f_1}\right)\delta\psi' \Dot{S}_g + d_1S_e^2 + d_2S_e'^2 + d_3\Dot{S}_e^2+ \left(e_1-\dfrac{f_5^2}{4f_1} + \dfrac{1}{4}\left(\dfrac{f_4f_5}{f_1}\right)'\right)\delta\psi^2  + \left(e_2-\dfrac{f_4^2}{4f_1}\right)\delta\psi'^2\nonumber\\
&+ e_3 \Dot{\delta\psi}^2  + f_1 \Bigl[2S_h\Bigl(H_0' - \dfrac{2}{A}\Dot{S}_1 + \dfrac{f_2}{2f_1}\Dot{S}_g + \dfrac{f_4}{2f_1}\delta\psi'+ \dfrac{f_5}{2f_1}\delta\psi - \dfrac{Af_3}{4f_1}H_0\Bigr) - S_h^2\Bigr]\,,
\end{align}
from which we can obtain algebraic expressions for $H_0$ and $S_1$ by variation with respect to each function
\begin{align}
H_0 &= \dfrac{1}{2\Bigl(a_1 - \dfrac{A^2f_3^2}{16 f_1} - \dfrac{(Af_3)'}{4}\Bigr)}\Bigl[\left(a_2 + \dfrac{Af_3f_5}{4f_1}\right)\delta\psi+ \left(a_3 + \dfrac{Af_3f_4}{4f_1}\right)\delta\psi' + \left(a_4 + \dfrac{Af_3f_2}{4f_1}\right)\Dot{S}_g\nonumber\\
&- \dfrac{Af_3}{2}S_h - 2 f_1' S_h'\Bigr],\\
S_1 &= \dfrac{1}{2b_1}\left[b_2 S_g + b_3\Dot{\delta\psi} +  b_4 S_g' + \dfrac{4f_1}{A}\Dot{S}_h \right].
\end{align}
Introducing them in our Lagrangian we get 
\begin{align}
\label{L_sca9}
\mathcal{L}_{\mathrm{sca}}^{(2)} &= \alpha_1 \Dot{S}_g^2 + \beta_1 S_g'^2 + \gamma_1 S_g^2 + \alpha_2 \Dot{S}_h^2 + \beta_2 S_h'^2 + \gamma_2 S_h^2+ \alpha_3 \Dot{S}_e^2 + \beta_3 S_e'^2 + \gamma_3 S_e^2 + \alpha_4 \Dot{\delta\psi}^2 + \beta_4 \delta\psi'^2 \nonumber\\
& + \gamma_4 \delta\psi^2 + \sigma_1 S_h\delta\psi + \sigma_2 S_h \delta\psi' + \sigma_3 S_h' \delta\psi' + \sigma_4 S_h' \Dot{S}_g + \sigma_5 S_g \Dot{S}_h + \sigma_6 S_g \Dot{\delta\psi} + \sigma_7 S_g' \Dot{\delta\psi} + \sigma_8 \Dot{S}_h \Dot{\delta\psi}\,,
\end{align}
where $\alpha_i$, $\beta_i$, $\gamma_i$ and $\sigma_i$ are coefficients that depend on $r$ only. These coefficients are too large to be written but can be shared upon request by email. 

Before closing this section, we will go back to the Lagrangian which decoupled from the rest. Using the following rescaling,
\begin{align}
S_e(t,r) = \sqrt{8(\lambda + 2(e^2r^2+ 2\kappa \lambda)\bar{\psi}^2)}S_m(t,r)\,.
\end{align}
This part of the action, upon transforming to the tortoise coordinate, takes the following form 
\begin{align}
\label{eq:ghost}
\int {\mathrm  d}t {\mathrm  d}r_* \Bigl[-\Bigl(\frac{\partial S_m}{\partial t}\Bigr)^2+\Bigl(\frac{\partial S_m}{\partial r_*}\Bigr)^2+V_m S_m^2\Bigr]\,,
\end{align}
where $V_m$ is given explicitly in the Appendix \ref{Appendix:Vm}. The presence of a global negative sign suggests the potential for ghostlike behavior. But, in the linear regime, with the fields considered in this action-namely the Maxwell field and the complex scalar field-the term $S_m$ is decoupled. Consequently, we can redefine $S_m \rightarrow i S_m$ without issue, thereby recovering a standard form of the action. We will see in section \ref{section:ghost} that no ghost is present.

\section{Second-order action for dipole perturbations, $\ell = 1$}

For dipole perturbations we expect to have three equations representing the contribution from the Maxwell field and scalar field. As we have seen in the vector case, we have $Y_1^0 \propto \cos{\theta}$ which implies
\begin{align*}
    h_{22} &= (K(t, r) -G(t, r))\cos{\theta},\quad
    h_{33} = (K(t, r) -G(t, r))\cos{\theta}\sin^2{\theta},\quad
    h_{23} = 0.
\end{align*}
We use the angular part of the diffeomorphism to fix $G=K$, making $h_{ij}=0$ and the radial diffeomorphism to fix $\delta\psi=0$ as well. The steps are similar to the general case with  $\delta\psi=0$ and $\lambda=2$. After some integration by parts we find the same action as (\ref{L_sca9}) with $\delta\psi=0$ and $\lambda=2$, namely

\begin{align}
\label{L_sca_l1_1}
\mathcal{L}_{\mathrm{sca}}^{(2)} &= \alpha_1 \Dot{S}_g^2 + \beta_1 S_g'^2 + \gamma_1 S_g^2 + \alpha_2 \Dot{S}_h^2 + \beta_2 S_h'^2 + \gamma_2 S_h^2+ \alpha_3 \Dot{S}_e^2 + \beta_3 S_e'^2 + \gamma_3 S_e^2 + \sigma_4 S_h' \Dot{S}_g  + \sigma_5 S_g \Dot{S}_h \,.
\end{align}
Similarly to the general case, the perturbation $S_m$ decouples and shows the same ghost-like potential behavior. 

\section{Second-order action for monopole perturbations, $\ell = 0$}
\label{section:ghost}

In the case $\ell=0$ we will have only two perturbation equations related to the scalar field degrees of freedom. Notice that for $\ell=0$ we have $Y_\ell^m = Y_0^0 = \mathrm{const}$, therefore, without using any gauge freedom we have $\delta A_i = h_{0i} = h_{1i} = 0$. Also we can use the radial diffeomorphism to fix $K=0$ and the time diffeomorphism for $H_1=0$, leaving a residual gauge $H_0 \rightarrow H_0 - F(t)$. The Lagrangian we have obtained is similar to (\ref{L_sca1}) with $\alpha=\beta=H_1=0$ and taking $\lambda=0\hspace{1mm} (\ell=0)$. We proceed in the very similar way as in the previous cases. For that we define $S_e$ as in eq.(\ref{elec_Se}) and obtain algebraic equation for $A_0$ (\ref{elec_Se_1}) and $A_1$ (\ref{elec_Se_111}). The Lagrangian has the following structure
\begin{align}
\label{L_sca1_l0_1}
\mathcal{L}_{\mathrm{sca,}l=0}^{(2)} &= a_1 H_2^2 + H_2[a_2\delta\psi + a_3 H_0' + a_4\delta\psi'] + b_1\delta\psi^2+ b_2\delta\psi'^2 + b_3 \Dot{\delta\psi}^2 + b_4\delta\psi H_0' + c_1S_e^2 + c_2S_e'^2+ c_3 \Dot{S}_e^2,
\end{align}
where the coefficients are not related to coefficients of Appendix (\ref{appendix:coeff}). As we have obtained in the previous cases, $S_e$ decouples. Also $H_0$ appears only as $H_0'$. By variation of the action with respect to $H_2$ we get the following expression
\begin{equation}\label{H_0'_l0}
    H_0' = -\dfrac{1}{a_3}\left[2a_1H_2 + a_2\delta\psi + a_4 \delta\psi' \right],
\end{equation}
which we can integrate and get an algebraic expression for $H_0$ plus a constant of integration $f(t)$ that we can eliminate using the residual gauge freedom. The new lagrangian obtained can be used to eliminate $H_2$ because it appears as a Lagrange multiplier. After some more integration by parts, we find
\begin{align}
\label{L_sca1_l0_2}
\mathcal{L}_{\mathrm{sca,}\ell=0}^{(2)} &= \alpha_1\delta\psi^2 + \alpha_2\delta\psi'^2 + \alpha_3 \Dot{\delta\psi}^2 + \beta_1S_e^2+ \beta_2S_e'^2 + \beta_3 \Dot{S}_e^2,
\end{align}
Finally, we use the following rescaling
\begin{align}
S_e(t, r) = 4er\Bar{\psi}S_m(t, r)\,,\quad
\delta\psi(t, r) = \dfrac{S_\psi(t, r)}{2r}\,,
\end{align}
and working with the tortoise coordinate, we obtain
\begin{align}
\label{eq:Pl0}
{S}_{\mathrm{sca}, \ell=0}^{(2)}=& \int {\mathrm  d}t {\mathrm  d}r_* \Bigl[\Bigl(\frac{\partial S_\psi}{\partial t}\Bigr)^2-\Bigl(\frac{\partial S_\psi}{\partial r_*}\Bigr)^2-V_{\psi} S_\psi^2 -\Bigl(\frac{\partial S_m}{\partial t}\Bigr)^2+\Bigl(\frac{\partial S_m}{\partial r_*}\Bigr)^2+{V_{m,\ell=0}} S_m^2\Bigr]\,,
\end{align}
where
\begin{flalign}
\label{eq:ghostl0}
V_{m,\ell=0}(r) &= A\biggl[-m^2 - \dfrac{1}{r^2} + 2\bar{\psi}^2\biggl(m^2 + \dfrac{e^2}{1+4\kappa\bar{\psi}^2}\biggr) + \dfrac{B}{r^2}\bigg(3+\dfrac{2r\bar{\psi}'(2\bar{\psi} + r\bar{\psi}')}{\bar{\psi}^2}\biggr)\biggr] 
+ e^2\bar{A}^2 &&\nonumber\\&
+ B\biggl(1+2\kappa + 4\kappa\bar{\psi}^2\biggr)\bar{A}'^2,
\end{flalign}
\begin{flalign}
\label{Vpsi}
V_{\psi}(r) &= A\biggl[m^2 + \dfrac{1}{r^2} - \dfrac{B}{r^2} - 2m^2\bar{\psi}^2 - 4\bar{\psi}'^2 + 8m^2r\bar{\psi}\bar{\psi}'(1+r\bar{\psi}\bar{\psi}')\biggr] + 3e^2\bar{A}^2
+\dfrac{B \bar{A}'^2}{1+4\kappa \bar{\psi}^2}\biggl[-1 &&\nonumber\\&\quad 
-2\kappa - 8\kappa\bar{\psi}^2(1-3\kappa + 2\kappa\bar{\psi}^2) - 16r\kappa\bar{\psi}(1+4\kappa\bar{\psi}^2)\bar{\psi}'+ 4r^2(1+4\kappa\bar{\psi}^2)^2\bar{\psi}'^2\biggr]\,.
\end{flalign}

We recover the possible ghost-like mode encountered earlier. However, one can analyze the monopole perturbation more straightforwardly in a gauge where only the complex scalar field is perturbed, while the metric and Maxwell field remain unchanged. In this context, for any generic background the perturbation is given by 
\begin{align} 
S &= 4 \int \sqrt{-\bar{g}}\Bigl[\frac{1}{A}\Bigl((\dot{\delta\psi_R})^2+(\dot{\delta\psi_I})^2\Bigr)-B \Bigl(({\delta\psi_R'})^2+({\delta\psi_I'})^2\Bigr)+2e\frac{\bar A}{A}\Bigl(\delta\psi_I\dot{\delta \psi_R}-\delta\psi_R\dot{ \delta \psi_I}\Bigr)\nonumber\\
&\quad +\Bigl(\frac{e^2 \bar A^2}{A}-m^2-\kappa \bar F^2\Bigr)\Bigl(({\delta\psi_R})^2+({\delta\psi_I})^2\Bigr)\Bigr]\,,
\end{align} 
with the definition $\delta \psi=\delta\psi_R+i\delta\psi_I$. This clearly shows the absence of any ghost. We therefore conclude that the ghost previously encountered is a consequence of the chosen set of variables. Performing a final change of variable, $S_m\rightarrow i S_m$, is thus justified and restores the action to a form in which the perturbations are ghost-free, as expected.

\section{Quasinormal modes and quasibound states}

We numerically analyzed the stability of vector perturbations for $\ell \geq 1$ and scalar perturbations for $\ell=0$, as well as the decoupled perturbation for $\ell=2$. In all cases, the potentials were found to be positive and approach zero at spatial infinity, indicating the existence of quasinormal modes and a stable background solution. However, in the scalar case with $\ell=0$, the potential can become slightly negative and nonzero at infinity. This behavior suggests the presence of quasibound states associated with the massive scalar field in the background solution.

Once the equations are obtained, all perturbations are expressed in the form $F(t, r)=f(r) e^{-i \omega t}$, where $\omega$ is the complex frequency governing the stability of the black hole. If the imaginary part of $\omega$ is negative, the black hole is stable. The real part of $\omega$ corresponds to the time periodicity of each mode.

To solve the problem, appropriate boundary conditions must be imposed. At the horizon, we require an ingoing mode for all cases. At spatial infinity, we impose an outgoing mode for quasinormal modes (QNM) and a perturbation for quasibound states (QBS) which tends to zero. 

All perturbations reduce to a single or coupled second-order linear ordinary differential equation with coefficients determined by the background fields. These equations depend on $\omega$, and only a discrete, infinite spectrum satisfies the boundary conditions, corresponding to QNM and QBS. In this work, we focus exclusively on the frequency with the smallest absolute value in the imaginary part, known as the fundamental mode.

Contrary to the background analysis, we will not use the Chebyshev nodes of the second kind
(\ref{eq:secondkind}) because of diverging behavior on the boundaries. For that we exclude the two boundaries from our rescaled $r$ coordinate and use therefore Chebyshev nodes of the first kind
\begin{align}
z_\alpha = \dfrac{1}{2}\biggl(1+\cos{\left(\dfrac{2\alpha+1}{2N+2}\pi\right)}\biggr)\,, \quad \alpha=0,1,...,N
\end{align}
Once the equation is discretized at the collocation points, the problem reduces to a matrix form, where the matrix depends on the frequency $\omega$. A non-trivial solution is obtained by finding the zeros of the determinant of this matrix. However, for large $N$ (the number of collocation points), numerical errors can significantly affect the analysis.

To address this issue, we search for values of $\omega$ that correspond to zero eigenvalues of the matrix. For each $N$, this calculation can be performed, and modes are selected based on their stability as $N$ increases, ensuring convergence by varying the matrix size.

\begin{figure}
    \centering
    \includegraphics[width=0.5\linewidth]{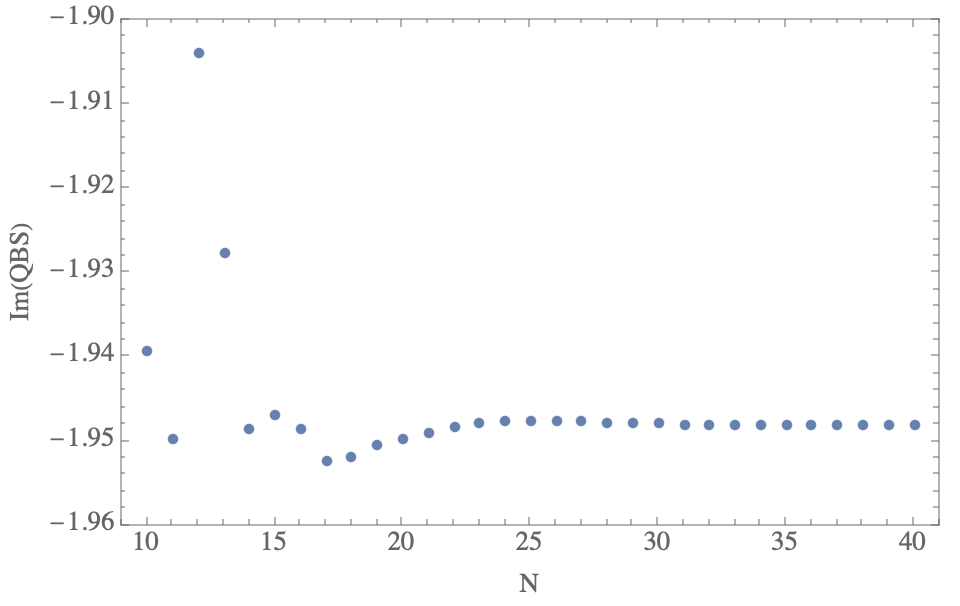}
    \caption{Convergence test of the fundamental mode associated to the perturbation $S_\psi$ as described in (\ref{eq:Pl0}), using the parameters $e=m=\kappa=1$, $\mu=0.95$ and  $y_H=0.5$. The boundary conditions are set to select the QBS. The real part is zero.} 
    \label{fig:convergence}
\end{figure}

For instance, in the case of scalar perturbations with $\ell=0$ and parameters $e=m=\kappa=1$, $\mu=0.95$, and $y_H=0.5$, we see in Figure (\ref{fig:convergence}) the evolution of the fundamental mode of the perturbation associated with the potential $V_\psi$ as a function of the parameter $N$, i.e. the number of points in our discretization. It is evident that for $N \geq 25$, the mode has stabilized and achieved convergence.

For all the background solutions we have analyzed, both quasinormal modes and quasibound states exhibit a negative imaginary part, indicating the stability of the black hole.

\begin{figure}
    \centering
    \includegraphics[width=0.45\linewidth]{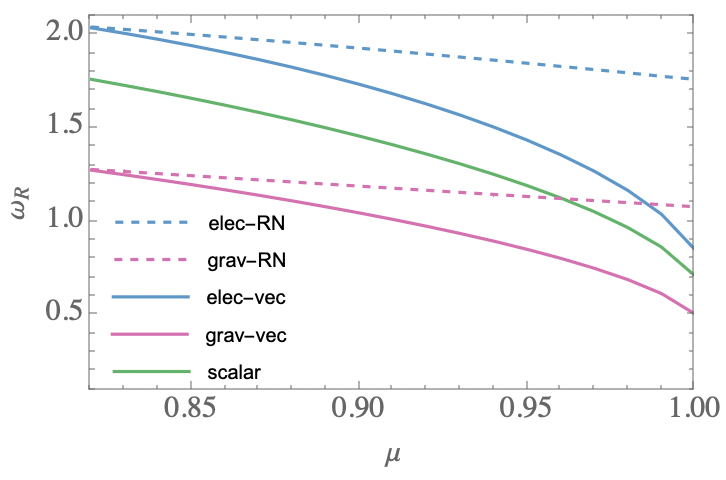}
    \includegraphics[width=0.48\linewidth]{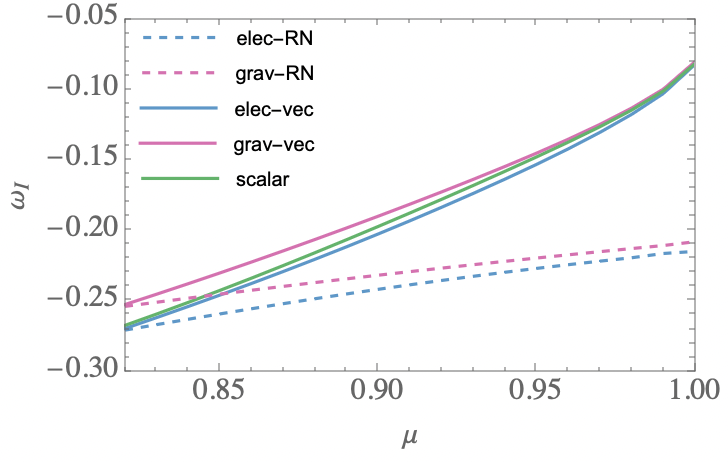}
    \caption{Real and imaginary part of the QNMs for $\ell=2$, $r_H=0.4, e= m = 1$ and $\kappa=1$ with $0.82\leq \mu\leq 1$ In green we show the QNMs associated to the perturbation (\ref{eq:ghost}) which correspond to the scalar perturbation, while the blue and red correspond to the 2 vector perturbations (\ref{accion_vec_final}). Finally, as a reference, we have included for the same values of $\mu$ and $r_H$, the corresponding QNMs for a RN background for $\ell =2$. We observe that as we approach the limit of existence of hairy BHs corresponding to the smallest value of $\mu$, we recover the QNMs of the RN black hole, showing continuity at the onset of existence of hairy BHs.} 
    \label{fig:QNM}
\end{figure}

As an example, Figure {(\ref{fig:QNM})} illustrates the QNMs for $\ell=2$, including both the vector perturbations described by Eq. (\ref{accion_vec_final}) and the decoupled scalar mode given by Eq. (\ref{eq:ghost}). For the background solution, we consider $r_H=0.4, e = m=1$, and $\kappa=1$, while $\mu$ varies from $0.82$, corresponding to the lowest value in which we found hairy black hole solutions, to $\mu=1$, representing the maximum value, known as the "warm hole". These modes are compared to the QNMs of the Reissner-Nordström (RN) black hole \cite{Moncrief:1974gw,Moncrief:1974ng}for $\ell=2$, which corresponds to the trivial solution with $\psi=0$.
In the limit $\mu=0.82$, as shown in Figure (\ref{fig:background}), the solution converges to the Reissner-Nordström background, and the quasinormal modes (QNMs) exhibit similar behavior. Indeed, for this set of parameters, and $\mu=0.82$, the solution is close to the onset of existence of hairy BHs. Conversely, as $\mu$ increases, the hairy black hole deviates further from the Reissner-Nordström solution, leading to increasingly distinct QNMs. Notably, the perturbation exhibits a smaller imaginary part compared to the Reissner-Nordström case, indicating that the perturbation persists longer around the black hole before dissipating. Surprisingly, for the largest values of $\mu$, the vector and scalar perturbations appear to share similar imaginary part of the QNMs. 

The QBSs can be studied in a similar form except that boundary conditions are modified. 

As an example, we have studied the monopole perturbation \eqref{Vpsi} for different values of $\mu$, with the coefficients fixed to $e = m = \kappa = 1$ and $y_H = 0.4$. QBSs reported in the literature are typically long-lived modes and thus have a small imaginary part in their frequency. In reality, however, this depends on the mass or, equivalently, on the value of the potential at infinity. In our case, \eqref{Vpsi} yields
\begin{align}
V(r=\infty)=m^2+3e^2\mu^2=1+3\mu^2
\end{align}
According to \cite{Detweiler:1980uk,Dolan:2007mj,Zouros:1979iw,Rosa:2011my,Huang:2020pga}, QBSs have a small imaginary part when $V(r \to \infty)$ is small. Therefore, even if in our case the imaginary part is relatively large, this is not inconsistent with the literature. For smaller values of $\mu$, we observe a kind of transition in which the real part of the modes is almost zero. These correspond to decaying modes with very small oscillations. Within our numerical precision, the modes remain oscillatory and thus correspond to genuine perturbations.
\begin{figure}
    \centering
    \includegraphics[width=0.5\linewidth]{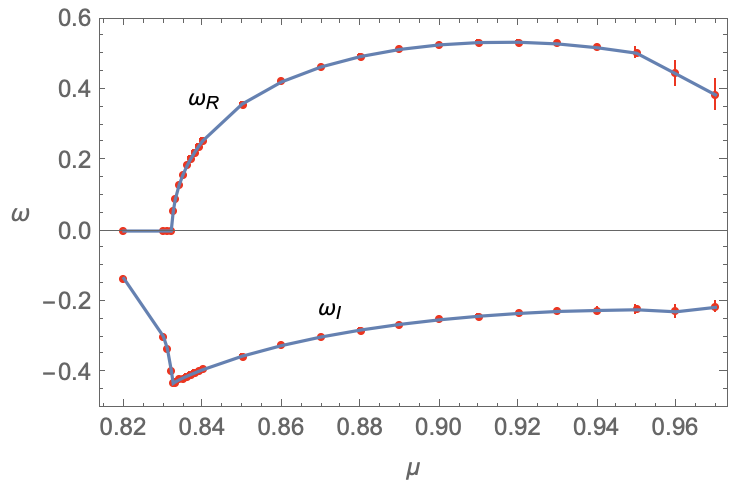}
    \caption{Real and imaginary parts of the QBSs for $\ell = 0$, $e = m = 1$, and $\kappa = 1$ with $r_H = 0.4$. The modes are shown as a function of $\mu$, the Maxwell potential at infinity.} 
    \label{fig:QBS}
\end{figure}
All QBSs studied here have a negative imaginary frequency, and therefore the spacetime is stable under these perturbations.

As discussed in Sec.~(\ref{existence}), the existence of a hairy black hole is constrained by the condition $e^2\mu^2 \leq m^2$. In this analysis, we have assumed $m/e = 1$, which restricts $\mu \leq 1$. To explore extremal solutions, we set $\mu = 1$ while keeping $m=e=\kappa = 1$. Unlike the previous case, the size of the horizon $r_H$ serves as the variable in this scenario. These parameters define what is referred to as a warm hole, an extremal solution where the temperature remains nonzero.  

\begin{figure}
    \centering
    \includegraphics[width=0.45\linewidth]{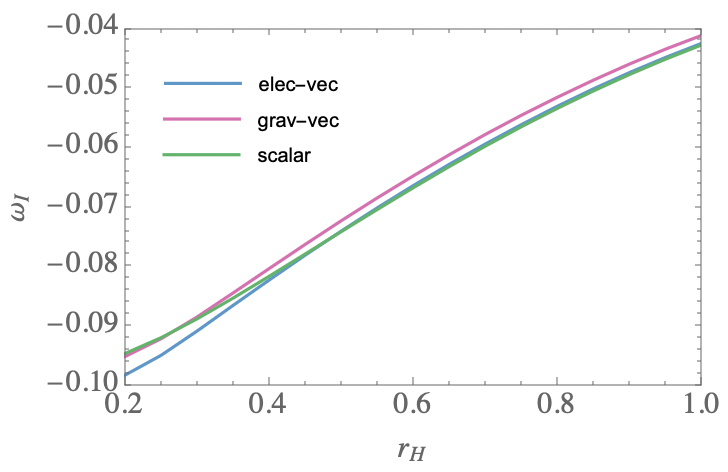}
    \caption{Imaginary part of the QNMs for $\ell=2$, $\mu=1, e=m = 1$ and $\kappa=1$ with $0.2\leq r_H\leq 1$. The parameters have been chosen such that we have an extremal solution, known as warm hole, for different size of the horizon.} 
    \label{fig:extremal}
\end{figure}

As illustrated in Figure~(\ref{fig:extremal}), the quasinormal modes (QNMs) differ for the three types of perturbations studied earlier. For larger extremal solutions, the perturbations decay more slowly compared to smaller black holes, similarly to standard black holes such as the Reissner-Nordström solution. Although the QNMs appear to be different within the limits of our precision, the scalar mode exhibits similar behavior to the gravitational mode for small black holes and to the electromagnetic mode for large black holes.

\begin{figure}
\centering
\includegraphics[width=0.45\linewidth]{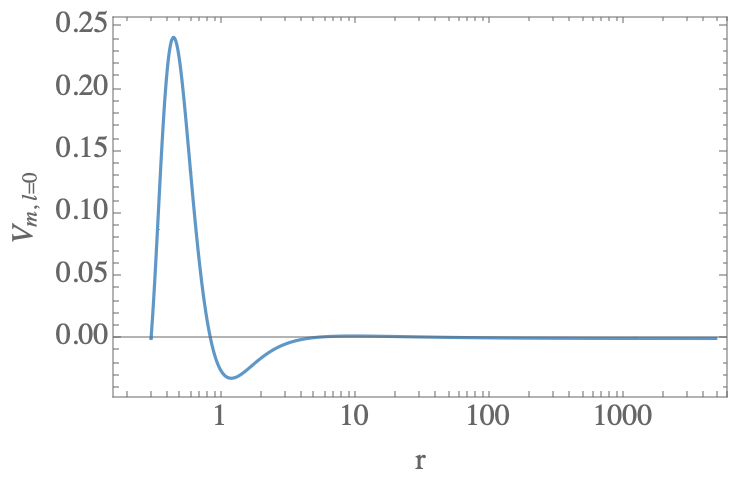}
\includegraphics[width=0.48\linewidth]{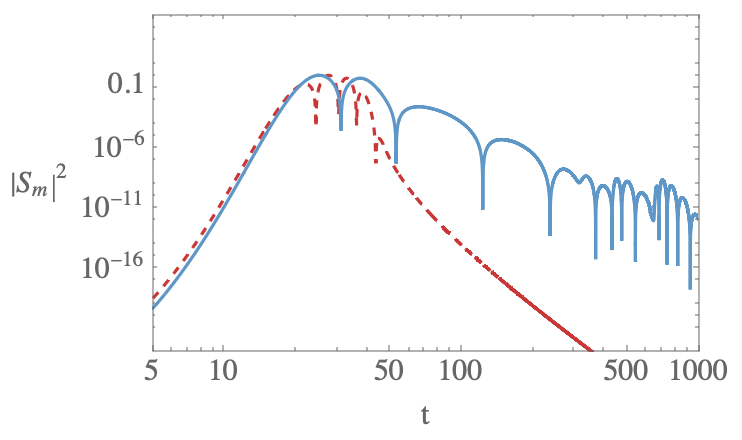}
\caption{Above: Potential (\ref{eq:ghostl0}) for $m=1$, $\kappa=0.7$, $y_H=0.3$, $\mu=0.9$, and $e=m/\mu$ (extremal case). The potential is negative.  
Below: Time evolution of the perturbation $S_m$ for this black hole, with $\ell=0$ at $r_*=5$, shown in blue. For comparison, the evolution of the perturbation for the Regge-Wheeler potential with $l=2$ is shown in red.} 
    \label{fig:stability}
\end{figure}

Finally, we have encountered situations where the potential is slightly negative, for example, for $m=1$, $\kappa=0.7$, $y_H=0.3$, $\mu=0.9$, and $e=m/\mu$, which corresponds to an extremal case. For this background, the potential of the perturbation described by (\ref{eq:ghostl0}) is negative, as shown in Figure (\ref{fig:stability}). 

Due to the numerical background solution, and therefore inaccuracies, we were unable to find the QNMs with sufficient precision. 

It is important to note that in this case the potential is negative at infinity,
\begin{align}
V(r=\infty)=-m^2+e^2\mu^2
\end{align}
Even so, it can be conveniently redefined by introducing
\begin{align}
\bar{\omega}^2=\omega^2+m^2-e^2\mu^2
\end{align}
which transforms the perturbation equation into
\begin{align}
    &\frac{d^2}{dr_*^2}S_m(r)+(\bar\omega^2-W(r))S_m(r)=0\,,\quad W(r)=V_{m,\ell=0}(r)+m^2-e^2\mu^2~.
\end{align}
The redefined potential no longer satisfies the standard condition $W(r_H) = 0$, yet an ingoing mode of the form
\begin{align}
    e^{-i\sqrt{\omega^2+m^2-e^2\mu^2}r_*}
\end{align}
can still be found near the horizon. For this reason, we do not expect such a mode to be particularly special. Unfortunately, the background solution did not provide sufficient accuracy to determine the QNMs. Therefore, to deduce stability, we analyzed the wave equation in null coordinates, $u = t - r_*$ and $v = t + r_*$, transforming the equation into  

\begin{align}
4(S_m)_{,uv} + V_{m,\ell=0} S_m = 0.
\end{align}  
This equation can be solved numerically, as outlined in \cite{Gundlach:1993tp}. As shown in Figure (\ref{fig:stability}), the black hole is stable.

The perturbation shows a longer ringing, likely due to numerical approximations. However, we observe a longer periodicity, which suggests that the real part of the QNMs is smaller than that of the Regge-Wheeler potential, $|\omega_R| < 2.49$, corresponding to the RW case with $r_s = 0.3$. Similarly, the slower decay of the perturbation indicates that $-0.59 < \omega_I < 0$, which is consistent with the fundamental mode found using the spectral method. Within this method, we observed that the fundamental mode did not converge with changes to the matrix size, but it appears to approach a value such that $|\omega_R| < 0.05$ and $-0.005 < \omega_I$. 

\section{Conclusions}

In this paper, we have investigated the stability of a charged hairy black hole, which exhibits intriguing properties such as the absence of a Cauchy horizon and the presence of a warm extremal solution. We derived the perturbation equations around a generic spherically symmetric background. In the vector sector, we identified two independent perturbations, while the scalar sector features four.  

The vector perturbations were analyzed in detail, demonstrating that they decay over time, thereby confirming the stability of the black hole under such perturbations. This analysis employed the $S$-deformation technique for all $\ell \geq 1$.  

In the scalar sector, we found that one degree of freedom decouples from the others. Although this mode appears with the "wrong" sign in the action—suggesting a potential instability, but the sign is irrelevant in the linear regime due to the decoupling. In the monopole case, we have shown using another gauge that no ghost is present. We conclude therefore that the wrong sign is due to the set of variables we have used and not a physical ghost. That means that a simple change of variable such as $S_m\rightarrow i S_m$ is justified rendering the action ghost free. The remaining three scalar modes are coupled, and their complete stability analysis remains an open problem. Specifically, we identified the perturbation equations for these modes, but their stability will be addressed in future work. 

Additionally, we studied the stability of the quasinormal modes (QNMs) and quasibound states (QBSs) associated with the decoupled mode. Despite the associated potential possibly being negative, this mode was found to be stable.

Finally, we explored the extremal case, corresponding to the maximally charged black hole. These solutions were shown to be stable, and we calculated their QNMs or their time domain evolution. Among the large number of black hole solutions we have studied, we found them to be stable.

\section*{Acknowledgments}
This work has been supported by ANID FONDECYT Regular No. 1220965 (Chile). The work of A.G.A was also supported by Beca ANID No. 22240063 (Chile).

\appendix

\section{Coefficients of Lagrangian (\ref{L_sca1})}
\label{appendix:coeff}
\allowdisplaybreaks
\begin{align}
a_1&=\dfrac{Q \Bar{A}'}{8} + \dfrac{A}{4}\mathcal{E}_A + C\,, \quad
a_2=\dfrac{\lambda}{2r}\sqrt{\dfrac{A}{B}}\left(2B + rB'\right)       \,,\quad
a_3=A\dfrac{\partial \mathcal{E}_A}{\partial \Bar{\psi}}          \,, \quad
a_4=\dfrac{4C}{\Bar{A}}     \,, \quad
a_5=\lambda \sqrt{AB}    \,,  \\
a_6&=\dfrac{J}{2} \,,\quad
a_7=\dfrac{Q}{2}\,, \quad
b_1=\dfrac{\lambda}{2}\sqrt{\dfrac{B}{A}} -\dfrac{B}{A}\left(4 C + B \mathcal{E}_B\right) \,,\quad
b_2=\dfrac{\lambda A'}{A} \sqrt{\dfrac{B}{A}}\,, \quad
b_3=-\lambda\sqrt{\dfrac{B}{A}}\,, \\
b_4&=b_3 \,,\quad
b_5=\dfrac{J}{A} \,,\quad
b_6=-\dfrac{8CB}{\Bar{A}} \,,\quad
c_1=\dfrac{1}{8}\left(Q \Bar{A}' + 2 B \mathcal{E}_B + 4 \sqrt{\dfrac{B}{A}}\left(A + r A' - 2r^2 A \Bar{\psi}'^2\right)\right) \,, \\
c_2&=-\dfrac{Q \Bar{A}'}{4} - \dfrac{\lambda}{2}\sqrt{\dfrac{A}{B}} + \dfrac{B\mathcal{E}_B}{2} + 2 C \,,\quad
c_3=\dfrac{4C}{\Bar{A}}\,, \quad
c_4=-\dfrac{\lambda}{2}\sqrt{AB}\left(\dfrac{A'}{A} + \dfrac{2}{r}\right) \,,\quad
c_5=\dfrac{8C}{\Bar{\psi}} - a_3 \,,\\
c_6&=-a_7 \,,\quad
c_7=r\sqrt{AB} \,,\quad
c_8=-2r\sqrt{\dfrac{B}{A}} \,,\quad
c_9=a_6 \,,\quad
c_{10}=\dfrac{\lambda}{\sqrt{AB}}\,,\\
d_1 &=\dfrac{\lambda \sqrt{AB}}{r^2}\left(1+\mathcal{E}_B B \sqrt{\dfrac{B}{A}}\right) \,,\quad
d_2=-\dfrac{\lambda Q}{r^2} \,,\quad
d_3=\dfrac{\lambda J}{r^2} \,,\\
d_4&=\dfrac{\lambda}{2r^2A}\Bigl(2\sqrt{\dfrac{A}{B}}\left(-1 + B -2r^2B\Bar{\psi}'^2\right) -4 C - A \mathcal{E}_A - B \mathcal{E}_B\Bigr)\,, \\
d_5&=-d_2 \,,\quad
d_6=\dfrac{\lambda}{2}\sqrt{\dfrac{B}{A}}\,, \quad
d_7=b_3 \,,\quad
d_8=d_6 \,,\quad
d_9=\dfrac{2\lambda}{r}\sqrt{\dfrac{B}{A}} \,,\quad
e_1=\dfrac{1}{2}\left(\dfrac{\partial \mathcal{E}_{\Bar{\psi}}}{\partial \Bar{\psi}} - \dfrac{\lambda J}{r^2 B \Bar{\psi}'}\right) \\
e_2&=\dfrac{16 C}{\Bar{A}\Bar{\psi}} \,,\quad
e_3=32r^2\kappa \sqrt{\dfrac{B}{A}}\Bar{\psi}\Bar{A}' \,,\quad
e_4=-4r^2\sqrt{AB}\,, \quad
e_5=\dfrac{4r^2}{\sqrt{AB}}\,,\\
f_1&=\dfrac{\lambda Q}{2r^2 B\Bar{A}'} + \dfrac{4C}{\Bar{A}^2}\,, \quad
f_2= - AB f_1\,, \quad
f_3=\dfrac{Q}{2\Bar{A}'}\,.
\end{align}

\section{Potential $V_m$}
\label{Appendix:Vm}
\begin{align}
&V_m(r) = \Bigl[2e^3(e^2r^2 + 2\kappa \lambda)\bar{A}^2\bar{\psi}^2(\lambda + 2(e^2r^2 + 2\kappa \lambda)\bar{A}^2) + 2 B\bar{\psi}^2(e^2r^2(1+2\kappa)+4\kappa^2\lambda + 4e^2r^2\kappa \bar{\psi}^2)(\lambda + 2(e^2r^2 \nonumber \\
&+ 2\kappa \lambda)\bar{\psi}^2)\bar{A}'^2 + \dfrac{A}{r^2(1+4\kappa\bar{\psi}^2)}(\lambda^3 - 2\lambda(e^2r^2(1+m^2r^2-3\lambda)+ 2\kappa(m^2r^2-3\lambda)\lambda)\bar{\psi}^2 -  4e^4r^4(1+m^2r^2)\bar{\psi}^4 \nonumber \\
&+ 4e^2r^2(r^2(3e^2 + m^2(1-6\kappa)) - 4\kappa)\lambda\bar{\psi}^4 + 16r^2\kappa
    (3e^2 - 2m^2\kappa)\lambda^2\bar{\psi}^4 + 48\kappa^2\lambda^3\bar{\psi}^4 + 12e^4r^4B\bar{\psi}^4 \nonumber \\
&+ 8(m^2r^2(e^4r^4(1 -2\kappa) + 
    + 4e^2r^2(1-2\kappa)\kappa\lambda - 8\kappa^3\lambda^2) + (e^2r^2 + 2\kappa\lambda)(e^4r^4 + 4\kappa^2\lambda^2 + 2e^2r^2\kappa
    (-1 + 2\lambda)) \nonumber \\
&+ 6e^4r^4\kappa B)\bar{\psi}^6 + 32m^2e^2r^4\kappa(e^2r^2 + 2\kappa\lambda)\bar{\psi}^8 + 4 r B \bar{\psi}
    (1+4\kappa\bar{\psi}^2)(\lambda(-e^2r^2 + 2\kappa\lambda)\nonumber \\
& + 4(e^2r^2 + \kappa\lambda)(e^2r^2 + 2\kappa\lambda)\bar{\psi}^2)\bar{\psi}' + 2r^2 (e^2r^2 + 2\kappa \lambda)B(1+4\kappa\bar{\psi}^2)(-\lambda + 4(e^2r^2 + 2\kappa\lambda)\bar{\psi}^2)\bar{\psi}'^2))\Bigr]\nonumber \\
&/
    (\lambda + 2(e^2r^2 + 2\kappa \lambda)\bar{\psi}^2)^2
\end{align}

\bibliographystyle{unsrt}   
\bibliography{biblio}     

\end{document}